\documentclass[a4paper,11pt]{article}
\usepackage{graphicx}
\usepackage{amsmath}
\usepackage{amsfonts}
\usepackage{amssymb}
\usepackage{epsfig}
\newcommand{\be}{\begin{equation}}
\newcommand{\ee}{\end{equation}}

\newcommand{\ba}{\begin{eqnarray}}
\newcommand{\ea}{\end{eqnarray}}

\begin{document}

\begin{flushright}%
ICCUB-09-338
\end{flushright}
\vskip   2.0 true cm

\begin{center}

{\LARGE \textbf{\bf The axion shield}} \\ [25 pt]

\textsc{A. Andrianov}\\
\textsl{V.A. Fock Institute of Physics,
Sankt-Petersburg State University,\\ 
Ulianovskaya 1,  198504 St. Petersburg, Russia,}\vskip0.2cm 
\textsc{D. Espriu}\footnote{On leave of absence from DECM and ICCUB, Universitat de Barcelona}
\\
\textsl{CERN, 1211 Geneva, Switzerland,}\vskip0.1cm
\textsc{and} \vskip0.1cm
\textsc{F. Mescia and A. Renau}\\
\textsl{Departament d'Estructura i Constituents de la
Mat\`eria and\\ Institut de Ci\`encies del Cosmos (ICCUB), Universitat de
Barcelona,\\
 Mart\'\i ~i Franqu\`es 1, 08028 Barcelona, Spain}

\vskip   1.5 true cm

\begin{abstract}
We investigate the propagation of a charged particle in a
spatially constant, but time dependent, pseudoscalar background.  
Physically this pseudoscalar background could be  provided
by a relic axion density. The background leads to an explicit breaking of 
Lorentz invariance; as a consequence
the process $p\to p \gamma$ is possible and the background 
acts as a shield against extremely energetic
cosmic rays, an effect somewhat similar to the GZK cut-off effect. 
The effect is model independent and can be computed exactly. 
The hypothetical detection of the photons radiated via this mechanism
would provide an indirect way of
verifying the cosmological relevance of axions.
\end{abstract}

\end{center}
\thispagestyle{empty}


\newpage
\setcounter{page}{1}

\section{Introduction}
Cold relic axions resulting from vacuum misalignment\cite{sikivie,raffelt} 
in the early universe 
is a popular and so far viable candidate to dark matter. If we assume that 
cold axions are the only contributors to the matter density of the universe
apart from ordinary baryonic matter its density must be\cite{wmap}
\be
\rho\simeq 10^{-30}{\rm g} {\rm cm}^{-3}\simeq 10^{-46} {\rm GeV}^4.
\ee
Of course dark matter is not uniformly distributed, its distribution 
traces that of visible matter (or rather the other way round). The 
galactic halo of dark matter (assumed to consist of axions) would correspond to a 
typical value for the density\cite{ggt}
\be
\rho_a\simeq 10^{-24}{\rm g} {\rm cm}^{-3}\simeq 10^{-40} {\rm GeV}^4
\ee
extending over a distance of 30  to 100 kpc in a galaxy such as
the Milky Way. Precise details of the density 
profile are not so important at this point.
Even in this case this is a rather low density; the axion background
provides a very diffuse concentration of pseudoscalar particles, but then 
the photon density is also low and yet it has an impact on ultra-high energy 
cosmic rays imposing the GZK cutoff \cite{GZK}. What are the consequences
of this difuse axion background on high-energy cosmic ray propagation? This is the
question we answer in this work.

The fact that the axion is a pseudoscalar, being the pseudo Goldstone
boson of the broken Peccei-Quinn symmetry\cite{PQ}, is quite relevant.
Its coupling to photons will take place through the anomaly
term; hence the coefficient is easily calculable once the axion model is known
\be
\Delta{\cal L}= g_{a\gamma\gamma} \frac{\alpha}{2\pi} \frac{a}{f_a}\tilde F F.
\ee
Two popular (and so far viable) axion models are the DFSZ\cite{DFSZ} and 
the KSVZ\cite{KSVZ} ones . In both models $g_{a\gamma\gamma}\simeq 1$. Here $a$ is
the axion field while $f_a$, the axion decay constant, is bound to be
of ${\cal O }(10^{11})$ GeV or more. As to the axion mass, $m_a$ it should lie
in the range $10^{-3} {\rm eV} > m_a > 10^{-6}$ eV.

The axion field is originally misaligned and in the process of relaxing to
the equilibrium configuration coherent oscillations with ${\bf q}=0$ are produced,
provided that the reheating temperature after inflation is below the 
Peccei-Quinn transition scale. In late 
times the axion field evolves according to
\be
a(t)= a_0 \cos m_a t.
\ee
Then, integrating by parts in the previous equation we get
\be
\Delta{\cal L}= -  g_{a\gamma\gamma} \frac{\alpha}{\pi} \frac{a_0}{f_a}
\sin (m_a t)\, \epsilon^{ijk} A_i F_{jk}\label{axionlag}
\ee
A particle travelling at the speed of light will see coherent regions with 
quasi-constant values of the axion background of a size ranging from the milimeter to
the meter, depending on the axion mass. Small as this size is, it is very many orders
of magnitude bigger that the wave length of a particle travelling with a
momentum ${\bf p}$ characteristic of a very high energy cosmic ray. For processes
with a high momentum transfer we can treat this slowly varying term 
as a constant and thus replace (\ref{axionlag}) by 
\be
\Delta{\cal L}= \frac14 \eta \epsilon^{ijk} A_i F_{jk},\label{liv}
\ee
where the ``constant'' $\eta$ will change sign with a period $1/m_a$. $\eta$ is obviously
related to the axion density. Numerically we expect
\be
\vert \eta\vert \simeq  g_{a\gamma\gamma} \frac{2\alpha}{\pi}  \frac{\sqrt{\rho_a}}{f_a}\simeq
10^{-24} - 10^{-25}{\rm eV},
\ee
for $f_a$ in the range $10^{11}-10^{12}$ GeV and where we have included
and additional factor of $\frac12$ to roughly account for the variability 
of the $\sin m_at$ term. 
Equation (\ref{liv}) can also be written as
\be
\Delta{\cal L} = \frac12 \eta_\alpha A_\beta \tilde F^{\alpha\beta},
\ee
with $\eta_\alpha=(\eta,0,0,0)$. The expression uses a covariant language 
but it is of course Lorentz invariance violating as $\eta_\alpha$ is a constant 
axial vector.

\section{Solving QED in a cold axion background}

Let us study the effects 
of explicit breaking of Lorentz invariance by means of a time-like 
constant axial vector. Consider electromagnetism in such a background
\be
{\mathcal L} = {\cal L}_{{\rm INV}} + {\cal L}_{{\rm LIV}}
\ee
\be
{\cal L}_{{\rm INV}} = 
-\,{\textstyle\frac14}\,F^{\alpha\beta}F_{\alpha\beta}
+ \bar\psi[i \not\! \partial  -e  \not\! A -m_e]\psi\qquad
{\cal L}_{{\rm LIV}} =
{\textstyle\frac12}\,m_{\,\gamma}^{2}\,A_\mu\,A^\mu
+\,{\textstyle\frac12}\,\eta_\alpha A_\beta\widetilde F^{\,\alpha\beta}
\ee
It will be useful for us to keep $m_\gamma > 0 $. It need not be a fundamental mass; but
an effective or induced in-medium mass. Otherwise
gauge invariance is manifest.

The equations of motion are
\be
\left\{g^{\,\lambda\nu}\left(k^2-m^2_\gamma\right) +
i \,\varepsilon^{\,\lambda\nu\alpha\beta}\,\eta_\alpha\,k_\beta\right\}
\tilde A_\lambda(k)=0.
\ee
\noindent

We can build two
 complex and space-like chiral polarization vectors 
$\varepsilon^{\mu}_{\pm}(k)$which satisfy the orthonormality relations
\be
-\,g_{\,\mu\nu}\;\varepsilon^{\mu\,\ast}_{\pm}(k)\,
\varepsilon^{\nu}_{\pm}(k)=1\qquad\quad
g_{\,\mu\nu}\,\varepsilon^{\mu\,\ast}_{\pm}(k)\,\varepsilon^{\nu}_{\mp}(k)=0
\ee
\noindent
In addition we have
\be
\varepsilon^\mu_T(k) \sim k^\mu\qquad \varepsilon^\mu_L(k)\sim k^2\eta^\mu
-k^\mu \eta\cdot k \ee
\noindent
They fulfill
\be
g_{\,\mu\nu}\;\varepsilon^{\mu\,\ast}_{\,A}(k)\,\varepsilon^{\nu}_{\,B}(k)\,=\,g_{\,AB}\qquad\quad
g^{\,AB}\,\varepsilon^{\mu\,\ast}_{\,A}(k)\,\varepsilon^{\nu}_{\,B}(k)\,=\,g^{\,\mu\nu}.
\ee
These polarizations correspond only approximately to the usual ones of QED. As a consequence
light propagation in an axion background may be subject to modifications. However, for visible light 
and even less for radiowaves, there is no marked separation of scales with respect to the axion
mass and therefore the time variation of the background cannot be treated adiabatically. The net effect
is expected to be zero or extremely  small \cite{boundrot}.
We refer the reader to \cite{AGS,AEGS} for additional details.

The polarization vectors of positive and negative chirality are
solutions of the vector field equations if and only if
\be
k^{\mu}_{\pm}=(\omega_{{\bf k}\,\pm} , {\bf k})\qquad
\omega_{{\bf k}\,\pm}=\displaystyle
\sqrt{{\bf k}^2+m_{\gamma}^{2}\pm\eta |{\bf k}|}\qquad
\varepsilon^{\mu}_{\pm}({\bf k},\eta)=\varepsilon^{\mu}_{\pm}(k_{\pm})
\quad
\left(\,k^{0}_{\pm} = \omega_{{\bf k}\pm}\,\right).\label{disprel}
\ee
In order to avoid problems with causality we want $k_{\pm}^2 > 0$. For
photons of a given chirality (negative if $\eta>0$, positive if $\eta<0$) 
this can be if and only if
\be
|{\bf k}|<\frac{m^{2}_{\gamma}}{\eta}\equiv\Lambda_{\,\gamma}.
\ee
In fact for $m_\gamma=0$ these photons cannot exist as physical asymptotic states.
If they are produced they will eventually decay (to three photons of like
chirality) in a cascade process that leads to a red-shift.

As is known to everyone the processes $e^- \to e^- \gamma$ or 
$\gamma \to e^+ e^-$ cannot occur in vacuum. However in the present situation,
because of the preferred reference frame, physics is different in 
different frames and for the latter process
\be
\omega_{\,{\bf k}\,\pm}=\displaystyle
\sqrt{{\bf k}^2+m_{\,\gamma}^{2}\pm\eta\,|{\bf k}|}
=\sqrt{{\bf p}^2+m_{e}^{2}} + \sqrt{({\bf p}-{\bf k})^2+m_{ e}^{2}},
\ee
as discussed in \cite{AGS,AEGS} it is possible for photons of the opposite chirality 
(positive if $\eta>0$, negative if $\eta <0$) if
\be
|{\bf k}|\ge\frac{4m_{\,e}^{2}}{\eta}\equiv k_{th}
\qquad (m_\gamma=0)
\ee

Here we shall be concerned with the related process $e \to e\gamma$
whose kinematics will be discussed in detail in the following sections.
In fact since we shall be concerned about the possible influence 
of the pseudoscalar background on cosmic ray propagation, which consist mostly of
protons, we shall
discuss the process $p\to p\gamma$. In this case there is a large hierarchy between the proton
momentum and the time variation of the background. We shall see below that
the process $p\to p\gamma$ is perfectly possible kinematically. This is a prompt process
that is not affected by a slight (in relative terms) time variation of the background. The
net effect shall not average to zero.

As for the value of $m_\gamma$ to use, the density of electrons in the universe
is expected to be at most of the order of $n_e=10^{-7}$ cm$^{-3}$. Photons pick up a mass
of the order of
\be
4\pi\alpha\frac{n_e}{m_e}.
\ee
This number is of the order of $10^{-15}$ eV, but we shall assume the more conservative
limit $10^{-18}$ eV, compatible with some 
astrophysical bounds in the $10^{-16}$ to $10^{-17}$ eV region \cite{bounds}. It is a very small
number, but it is non-zero for sure (it actually depends on the local density of electrons, of course).

\section{Kinematic constraints}

Having found out the different polarizations and dispersion relations in the  axion background
let us now turn to kinematical considerations. Let us consider the process
$p({\bf p}) \to p({\bf p}-{\bf k})\gamma({\bf k})$ 
with $p^\mu=(E,{\bf p}), p=\vert {\bf p} \vert$ and 
$k^\mu=(w_{\bf k},{\bf k}), k=\vert {\bf k} \vert$. 
Using (\ref{disprel}) energy conservation leads to 
\be
\sqrt{E^2 + k^2 - 2 pk \cos\theta} + \sqrt{k^2 + \eta k + m_\gamma^2} - E = 0.
\label{encor}
\ee
For simplicity we take the positive sign for the polarization, in the understanding that changing the sign
of $\eta$ amounts to exchanging positive and negative chiral polarizations for the photon.

Let us first consider the case $m_\gamma=0$. Then the above energy conservation equation reduces to
\be
k^2 (4E^2 - 4p^2\cos^2\theta - \eta^2 - 4p\eta\cos\theta) +k (4p^2\eta + 4m_p^2\eta)=0.
\ee
This equation has the trivial solution $k=0$, where nothing happens, and 
\be
k= \frac{4p^2\eta + 4m_p^2\eta}{-4E^2 + 4p^2\cos^2\theta +\eta^2 + 4p\eta\cos\theta}.
\ee
To find out the kinematical restrictions on $k$ we search for the extrema of the denominator. There is only
one  such extremum which is
is $\cos\theta=-\eta/2p$. Plugging this into the expression for $k$ one gets 
$k=-\eta$. Since $k$ has to be positive
this process is kinematically impossible unless $\eta <0$. Of course for $\eta=0$ it is also impossible as
this is the familiar vacuum case. So the process $p({\bf p})\to p({\bf p}-{\bf k})\gamma({\bf k})$ is actually
possible, for positive values of $\eta$ if the emitted photon is of negative chirality. On 
the other hand, if $\eta$ is negative, the process becomes possible only for positive chirality photons. Here
we shall take $\eta$ to be always negative, but it is trivial to recover the results for positive $\eta$ by
exchanging chiralities.

This extremal value (for negative $\eta$) is clearly providing the minimum value for $k$, 
$k_{min}=\vert \eta\vert$. There is no other extremal point, so the maximum value for $k$ has to lie at one
of the two values $\cos\theta=\pm 1$. It is not difficult to convince oneself (and quite intuitive) that
the maximum is found for $\cos\theta=1$ and it corresponds to
\be
k_{max}= \frac{E^2}{p-\frac{m_p^2}{\eta}},
\ee
so $k_{max}\simeq E$ for $E\gg m_p^2/|\eta|$ and $k_{max}\simeq |\eta|E^2/m_p^2$ for $E\ll m_p^2/|\eta|$.

Let us now consider the case $m_\gamma>0$. 
Conservation of energy
leads to the following two roots for $k$
\be
k_{\pm}=\frac{2E^2\eta-2p m_\gamma^2 \cos\theta-\eta m_\gamma^2 \mp
2E \sqrt{E^2\eta^2-4E^2m_\gamma^2+m_\gamma^4+4p^2m_\gamma^2\cos^2\theta
+2p\eta m_\gamma^2\cos\theta}}
{-4E^2+4p^2\cos^2\theta +4p\eta\cos\theta +\eta^2}.\label{sqrt}
\ee
The equations above are meaningful only for those values 
$\theta$ and $p$ providing a positive discriminant $\Delta$. From the
condition that $\Delta\ge 0$ we get
\be
\sin^2\theta\le \frac{\eta^2}{4p^2m_\gamma^2}\frac{1}{1+\frac{\eta}{4p}}
(p-p_-)(p-p_+),\label{sin}
\ee
where
\be
p_\pm= \frac{m_\gamma^2}{-\eta}\pm\frac{2m_p m_\gamma}{-\eta}\sqrt{1-\frac{\eta^2}{4m_\gamma^2}}.
\ee
Let us examine these two values $p_\pm$. Since $\eta$ is taken negative, it is clear
that $p_+ >0$ and $p_-<0$. Therefore for $\sin^2\theta$ to be a positive quantity we need
\be
p>p_+= p_{th}\simeq \frac{2m_\gamma m_p}{-\eta}.
\ee
This is the threshold value below which the process cannot take place kinematically.
For $p>>p_{th}$, 
\be
\sin^2\theta_{max} \to \frac{\eta^2}{4m_\gamma^2}.
\ee
$\theta_{max}$ is the value of $\theta$ where the bound in $\sin^2\theta$ of eq.(\ref{sin}) is saturated.
We see that photons are emitted in a rather narrow cone. This justifies {\em a posteriori}
the approximation $\cos\theta \simeq 1- \frac12 \sin^2\theta$ that has been used.

At $\theta_{max}$ the square root in (\ref{sqrt}) vanishes and $k_+ = k_-=k(\theta_{max})$ in (\ref{sqrt}). 
Then, keeping only the leading terms we get
\be
k(\theta_{max})\simeq \frac{2m_\gamma^2}{-\eta}(1 + 3 \frac{p
m_\gamma^2}{E^2\eta})\stackrel{p>>p_{th}}{\longrightarrow} \frac{2m_\gamma^2}{-\eta}
\ee
From eq.(\ref{sqrt}) we work out the value for $\theta=0$, which is the minimum value of $\theta$ from the 
bound in eq.(\ref{sin}).  
\be
k_{+}(0)\simeq \frac{E^2\eta -pm_\gamma^2 - E\sqrt{E^2\eta^2-4m_p^2m_\gamma^2+2p\eta m_\gamma^2}}{2p\eta - 2 m_p^2}
\stackrel{p>>p_{th}}{\longrightarrow} \frac{E^2}{p-\frac{m_p^2}{\eta}},
\ee
which is the same result obtained before, and
\be
k_{-}(0)\simeq \frac{E^2\eta -pm_\gamma^2 + E\sqrt{E^2\eta^2-4m_p^2m_\gamma^2+2p\eta m_\gamma^2}}{2p\eta - 2 m_p^2}
\stackrel{p>>p_{th}}{\longrightarrow}\frac{m_\gamma^2}{-\eta}
\label{tresh}
\ee
Now we notice that $k_-(0)<k(\theta_{max})<k_+(0)$. To show that 
$k_{min}=k_-(0)$ and $k_{max}=k_+(0)$ we have to study the derivative 
of $\theta$ 
versus $k$, namely we should have  
$d\cos\theta/dk<0$ for $k<k(\theta_{max})$ and 
$d\cos\theta/dk>0$ for $k>k(\theta_{max})$.

We isolate $\cos\theta$ from
the energy conservation relation (\ref{encor})
\be
\cos\theta = \frac{m_\gamma^2 +\eta k -2 E\sqrt{m_\gamma^2+k^2+\eta k}}{-2pk}.\label{cos}
\ee
and 
\be
\frac{d\cos\theta}{dk}= \frac{m_{\gamma}^2}{2 k^2 p} - 
\frac{E}{2 k^2 p} \frac{\eta k+ 2 m_{\gamma}^2}{\sqrt{\eta k + k^2 + m_{\gamma}^2}}
\stackrel{p>>p_{th}}{\longrightarrow} 
-\frac{\eta k+ 2 m_{\gamma}^2}{2 k^2 \sqrt{\eta k + k^2 + m_{\gamma}^2}}
\ee
For $k\to k(\theta_{max})$ $d\cos\theta/dk\to 0$ and is the only zero which of course
means that this value of $k$ corresponds to a minimum of $\cos \theta$ (i.e. to
a maximum of $\sin\theta$). On the other hand,  
for $k<k(\theta_{max})$, $d\cos\theta/dk<0$ and for
$k>k(\theta_{max})$, $d\cos\theta/dk>0$.
Then $k_{min}=k_-(0)$ and $k_{max}=k_+(0)$

We have said before that the actual cone of photon emission is quite narrow
for the values of $m_\gamma^2$ and $\eta$ relevant in the present discussion, but 
for the sake of completeness let us discuss the case where $m_\gamma$ is small.
We use again the expression (\ref{cos}) and
\be
\frac{dk}{d\cos\theta}= 
\frac{2pk^2\sqrt{m_\gamma^2+k^2+k\eta}}{m_\gamma^2\sqrt{m_\gamma^2+k^2+k\eta} -2Em_\gamma^2}.
\ee
The extremum conditions leads to $k=0$ or $m_\gamma^2+k^2+k\eta=0$. The first possibility is
unphysical since it is incompatible with (\ref{cos}). The second one gives
\be
k=\frac{-\eta\pm\eta\sqrt{1-\frac{4m_\gamma^2}{\eta^2}}}{2},
\ee
which is real only as long as $\eta^2 > 4 m_\gamma^2$. These expressions have a
smooth $m_\gamma^2\to 0$ limit which has already been discussed at the beginning of this section.

\section{Differential decay probability}

Here we shall compute the relevant matrix element and the differential
decay probability. The calculation is standard but in view of the
rather peculiar properties of Chern-Simons QED it is probably useful to
go into some details.
Since the process is possible only for one chirality of the photon
(depending on the sign of $\eta$) we shall not sum over the final polarization
of the radiated photon.

Using the standard Feynman rules we get 
\be
|M|^2=e^2\varepsilon^*_\mu\varepsilon_\nu
{\rm tr}\,[u(q)\bar{u}(q)\gamma^\mu u(p)\bar{u}(p)\gamma^\nu]
\ee
Summing and averaging over the final and initial proton helicities, respectively, 
and performing the trace we get
\be
\overline{|M|^2}=2e^2\left\{-p\cdot
k+\left[\varepsilon^*_\mu(k)\varepsilon_\nu(k)+\varepsilon_\mu(k)\varepsilon^*_\nu(k)\right]p^\mu
p^\nu\right\}= 2e^2\left(-p\cdot
k + 2\vert \varepsilon^\mu_-(k)p_\mu\vert^2\right),
\ee
where $\varepsilon^\mu_-({\bf k})$ is the polarization vector
defined in section 2.
Recall that \cite{AGS} 
\be
\varepsilon^\mu_-({\bf k})^*\varepsilon^\nu_-({\bf k})+\varepsilon^\mu_-({\bf k})
\varepsilon^\nu_-({\bf k})^* = - \frac{S^{\mu\nu}}{\eta^2 \vert{\bf k}\vert^2}
\ee
where
\be
S^{\mu\nu}=((\eta\cdot k)^2-\eta^2 k^2)g^{\mu\nu}
- (\eta\cdot k)(\eta^\mu k^\nu + \eta^\nu k^\mu) 
+ k^2\eta^\mu\eta^\nu + \eta^2 k^\mu k^\nu .
\ee
Then
\be
\overline{|M|^2}=2e^2\left(-p\cdot
k + \vert {\bf p}\vert^2 \sin^2\theta\right).
\ee
It is not difficult to see that in the kinematical conditions where
the process is possible $p\cdot k < 0$, thus guaranteeing positivity.

The differential decay width will be
\be
d\Gamma(Q)=(2\pi)^4\delta^{(4)}(q+k-p)\frac1{2E}\overline{|M|^2}dQ,
\ee
where $dQ$ refers to the final state phase space.
The final result is
\be
d\Gamma(Q)= \frac1{16\pi}\frac{|{\bf k}|}{|{\bf p}|}\frac{\overline{|M|^2}}{E\omega_{\bf k}}d|{\bf k}|
=\frac\alpha2\frac{|{\bf k}|}{|{\bf p}|}\frac1{E\omega_{\bf k}}(-p\cdot k+|{\bf p}|^2\sin^2\theta)d|{\bf k}|
\ee
after using the energy-momentum  conservation delta and performing the
integral over all the angles. In the previous expression $\sin\theta$ corresponds to
\be
\cos\theta=\frac{m_\gamma^2+\eta |{\bf k}|-2E\omega_{\bf k}}{-2|{\bf p}||{\bf k}|}.\label{cosinusk}
\ee

\section{Energy loss and radiation spectrum}

Since in this section no confusion with four-vectors is possible, we revert to the 
lighter notation
$p=\vert {\bf p} \vert, k=\vert {\bf k} \vert$ and $w=w_{\bf k}$.
The relevant quantity to compute is the following 
\be
\frac{dE}{dx} = - \frac{1}{v}\, \int d\Gamma(Q) w(Q).
\ee
Then, using the previous results,
\be
\frac{dE}{dx}= - \frac \alpha2\frac1{p^2}\int\,kdk [-\frac12(m_\gamma^2+\eta k)+p^2(1-\cos^2\theta)].
\ee
In the previous expression $\cos\theta$ is given by (\ref{cosinusk}).
Then the  l.h.s of this identity equals
\be
- \frac\alpha{8p^2}\int\frac{dk}k[-2\eta k^3-2m^2_\gamma k^2+4p^2k^2-\eta^2k^2
-m^4_\gamma-4E^2w^2-2m^2_\gamma\eta k+4Em^2_\gamma\omega+4\eta E\omega k].
\ee
Expanding $\omega=k\sqrt{1+\frac\eta k+\frac{m^2_\gamma}{k^2}}\approx k+\frac\eta2+\frac{m^2_\gamma}{2k}$, 
recalling the integration limits $\frac{m^2_\gamma}{-\eta}$ and  $\frac{\eta E^2}{p\eta-m^2_p}$ and neglecting at
this point the effective photon mass we get 
\be
\frac{dE}{dx}=-\frac\alpha{8p^2}\frac{\eta E^2}{(p\eta-m^2_p)^3}[-\frac23\eta^3E^4
+2\eta E^2 (p\eta-m^2_p)(m^2_p+\eta E-2\eta p-\eta\frac{m_p^2}{E}-\frac{\eta^2}{4}+\eta^2\frac{p}{E} )].
\ee
There are two relevant limits
\be
E \ll \frac{m_p^2}{|\eta|}\longrightarrow\frac{dE}{dx}=-\frac{\alpha\eta^2 E^2}{4m^2_p}.
\ee
\be
E\gg \frac{m_p^2}{|\eta|}\longrightarrow\frac{dE}{dx}=-\frac{\alpha|\eta|}{3}E
\ee
At this point it becomes obvious why we have bothered to keep the proton mass $m_p$ all along the calculation.

There are two key scales in this problem. One, discussed in sections 2 and 3, is the threshold energy
where the process $p \to p \gamma$ becomes kinematically possible in the presence of a pseudoscalar 
background represented by $\eta\neq 0$, namely
$E_{th}\simeq 2m_\gamma m_p/|\eta|$.
The other relevant scale is $m_p^2/|\eta|$ where there is a crossover: the energy loss
per unit length well below this energy is effectively proportional to $\eta^2$, 
while well above that scale is proportional to
$\eta$. Thus even if we are talking about very energetic particles the mass is a relevant parameter
when Lorentz violating interactions are present.

If $E \gg m_p^2/|\eta|$ the energy loss is given by
\be
E(x)= \exp{- \frac{\alpha|\eta|}{3}x}.\label{expo}
\ee
For $\eta$ in the range of values described in section 1, this would give a mean free path 
in the range ${\cal O}(1)$ to ${\cal O}(10)$ kpc. This would imply that cold axions act as a powerful shield
againts very energetic cosmic rays. Since $E_{th}\simeq 10^{15}$ eV the detection of cosmic rays above that
energy would in fact impose a rather stringent bound on the combination $\sqrt{\rho_a}/f_a$.

However, this is not so because even for the most energetic cosmics, just below the GZK cut-off 
of $10^{20}$ eV, we are
well below the cross-over scale $m_p^2/|\eta|$. In this regime the expression for $E(x)$ is
\be
E(x)= \frac{E(0)}{1+\frac{\alpha\eta^2}{4m_p^2} E(0) x}.
\ee
It is peculiar to see that for extremely large distances $E(x)\sim \frac{1}{x}$, 
independently of the energy
of the primary particle. However the distances where this behaviour could 
be seen are unphysically large because $\frac{\alpha\eta^2}{4m_p^2}$
is very small.

Thus the effect of the pseudoscalar background on the reach of cosmic rays is totally negligible 
for reasonable energies. Does this mean
that the effect is completely invisible? Not quite.

Let us assume for the sake of definiteness that $E(0)\simeq 10^{20}$ eV. Then, the photon emission
due to the axion background goes all the way up to $w_{max}\sim 10^{-2}$ eV (for $\sqrt{\rho_a}/f_a$ just 
in the currently allowed upper limit, and getting smaller as $\eta$ decreases). These energies corresponds to  
radio emission in the submillimetric range ($10^{-5}$ m or larger).

For electrons or positrons the numbers are different. The threshold energy for the radiation loss due
to the interaction with the axion background can be as low as 1 TeV (for $f_a \sim 10^{11}$ GeV). The
radiation emitted in this case would be with wavelengths characteristic of X-ray emission, or larger. 
 
It is not easy to say whether these photons could be detected at all, because even 
over cosmological distances the rate of radiation is very low for protons (higher for electrons), 
but the emission
of cosmic rays is sometimes copious and this could offset the low probability of radiation. 
The shape of the spectrum
is completely calculable by the formulae we provide here, and the radiation is emitted in a 
very narrow cone around the charged particle.
These two characteristics may be instrumental to detect it or, alternatively, to exclude its presence.
Excluding it with some confidence level would place an interesting bound on $\sqrt{\rho_a}/f_a$.
It should be mentioned that, being low energy, this radiation once is produced is not
affected by the pseudoscalar background as discussed in section 2.

\section{Conclusions and outlook}

In this work we have considered the effect on charged particles of 
a mildly (compared to the particle momentum) time dependent pseudoscalar background. A physical
situation worth exploring is the influence of a diffuse relic cold axion 
background on cosmic ray propagation. The effect is model independent, universal and can be 
computed unambiguously.

The effect is completely calculable in great detail because particle propagation
is governed by a modification of QED (Maxwell-Chern-Simons Electrodynamics) that is
exactly solvable. We have determined the kinematical constraints, the characteristics of the
emitted radiation and the rate of energy loss of charged particles moving in such a medium.
Some rather non-intuitive features appear and the results, we believe, are interesting per se.
We find that for protons the phenomenon may appear (depending on the value of
$\sqrt{\rho_a}/f_a$) at energies around $10^3$ TeV, but for those primaries  
with energies that survive the GZK cut-off the suppression
brought about by the interaction with this 'axion shield' is quite negligible, being only relevant
for extremely energetic particles. However, the 'bremmstrahlung' radiation may be
measurable. The characteristic energies and extremely strong colimation associated to given primaries
may be crucial for that. 

We have also investigated the relevance of this phenomena for electrons and positrons. In this case the
threshold energy is really around the corner (around 1 TeV, maybe even less as it depends on the photon
effective mass and on the value of $\sqrt{\rho_a}/f_a$) and the radiation could be as energetic as
X-rays.

An analogous phenomenon may present itself in a completely different context, namely heavy ion
collisions at high density and relatively low temperature. As emphasized recently \cite{ae} 
a phase where a parity breaking condensate is present may well appear even at moderate densities.
Once formed this condensate would be time dependent due to the expansion of the fireball produced
in the collision. Such a background would have rather non trivial effects on the propagation
of different particles. It is known to potentially influence the behaviour of scalars and 
pseudoscalars, dramatically changing the spectrum, but it would also influence, via a mechanism
similar to the one discussed here, photons and leptons. This effect is being investigated\cite{aex}.

\section*{Acknowledgements}

We acknowledge the financial support from projects FPA2007-66665, 
2009SGR502, Consolider CPAN CSD2007-00042 and EU ITN contract
MRTN-CT-2006-035482, ``FLAVIAnet".
A.A is also supported by grants RFBR 09-02-00073-a, 09-01-12179-ofi-m and program 
RNP2009-2.1.1/1575. A.R. thanks the Spanish Ministry of Education for a collaboration
grant.

\end{document}